\documentclass[letterpaper,10pt,conference]{ieeeconf}
\IEEEoverridecommandlockouts
\usepackage{cite}
\usepackage{amsmath,amssymb,amsfonts,bm,mathtools}
\allowdisplaybreaks[1]
\usepackage{graphicx,booktabs}
\usepackage[ruled,vlined,lined,boxed,linesnumbered,commentsnumbered]{algorithm2e}
\usepackage{algpseudocode}
\usepackage{balance}
\usepackage{tikz}
\usetikzlibrary{arrows.meta,fit}
\usepackage[utf8]{inputenc}

\newcommand{\E}{\mathbb E}
\newcommand{\N}{\mathcal N}
\newcommand{\IW}{\mathcal{IW}}
\newcommand{\Beta}{\operatorname{Beta}}
\newcommand{\Bern}{\operatorname{Bern}}

\newcommand{\sym}{\operatorname{sym}}

\newtheorem{theorem}{Theorem}
\newtheorem{proposition}{Proposition}

\newtheorem{assumption}{Assumption}

\def \cL {\mathcal{L}}
\def \cN {\mathcal{N}}
\def \cI {\mathcal{I}}
\def \by {\boldsymbol{y}}
\def \bh {\boldsymbol{h}}
\def \bz {\mathbf{z}}
\def \bl {\boldsymbol{\lambda}}
\def \bmu {\boldsymbol{\mu}}
\def \bTheta {\boldsymbol{\Theta}}
\def \bx {\boldsymbol{x}}
\def \bR {\boldsymbol{R}}
\def \bP {\boldsymbol{P}}

\title{\LARGE\bf
	Nonlinear Joint State and Noise Estimation under Packet Dropouts and Additive Measurement Corruption}

\author{Peng Sun$^{1}$, Ruoyu Wang$^{2}$, and Xue Luo$^{3,\sharp}$%
	\thanks{This work was supported by the National Natural Science Foundation
		of China under Grant 12271019 and the National Key R\&D Program of China
		under Grant 2022YFA1005103.}%
	\thanks{$^{1}$P. Sun is with the School of Mathematical Sciences, Beihang
		University, Beijing 102206, China. {\tt\small 1512530585@buaa.edu.cn}}%
	\thanks{$^{2}$R. Wang is with the School of Mathematical Sciences, Beihang
		University, Beijing 102206, China. {\tt\small WangRY@buaa.edu.cn}}%
	\thanks{$^{3}$X. Luo is with the School of Mathematical Sciences and the Key
		Laboratory of Mathematics, Informatics and Behavioral Semantics, Beihang
		University, Beijing 100191, China. {\tt\small xluo@buaa.edu.cn}}%
	\thanks{$^{\sharp}$Corresponding author.}}

\begin{document}
	\maketitle
	\thispagestyle{empty}
	\pagestyle{empty}
	\raggedbottom
	
\begin{abstract}
This paper investigates joint state and noise covariance estimation for nonlinear systems subject to dual reliability impairments: random packet dropouts and additive measurement corruption. This scenario poses challenging nonconjugate inference, as clean and corrupted measurements share an unknown baseline covariance $\boldsymbol{R}_k$ with fixed known corruption covariance $\boldsymbol{E}_k$. Packet reception is observable, whereas the authenticity of received measurements remains latent and must be inferred.  To resolve this intractable estimation problem, we cast joint posterior inference as a variational optimization problem and develop a conjugate variational inference adaptive Kalman filter with dual reliability (CVIAKF-DR). The proposed method retains analytical updates for the prediction covariance and two reliability probabilities, while adopting CVI to estimate the state and baseline measurement covariance. The covariance update combines the analytical contribution from clean measurements with Monte Carlo gradient estimates for corrupted measurements. Numerical results on a re-entry tracking benchmark demonstrate the superior accuracy of the proposed method in both state and noise covariance estimation. Remarkably, even built upon the linear Kalman filtering (KF) framework that is originally designed for linear systems, our method achieves more accurate estimation performance than dedicated nonlinear filtering algorithms based on the cubature Kalman filter (CKF).
	\end{abstract}
	
	\section{Introduction}
	Networked sensing systems are subject to two statistically distinct observational uncertainties: packet dropouts and measurement corruption. Packet loss reduces communication availability by discarding observations, whereas corruption produces erroneous yet successfully received measurements. When the baseline measurement covariance is unknown, ignoring this distinction may confound intrinsic sensor uncertainty with corruption-induced errors. We therefore distinguish observable packet reception from latent measurement authenticity and model them as two hierarchical reliability mechanisms. In addition, rather than assigning unrelated covariances to clean and corrupted measurements, we adopt a structured additive model in which their covariances are \(\boldsymbol{R}_k\) and \(\boldsymbol{R}_k+\boldsymbol{E}_k\), respectively, where \(\boldsymbol{R}_k\) is the unknown intrinsic measurement covariance and \(\boldsymbol{E}_k\) is a known positive-semidefinite corruption increment. This construction follows naturally when intrinsic measurement noise and additive corruption are independent, but the shared unknown \(\boldsymbol{R}_k\) destroys inverse-Wishart conjugacy and creates a nonconjugate covariance-inference problem.
	
	Existing related work can be broadly grouped into three directions. Packet-loss filtering models unreliable communication through Bernoulli arrivals \cite{1333199}, Markov channels \cite{Huang2007MarkovianLosses,You2011MeanSquare}, and, more recently, nonlinear Bayesian models with unknown loss probabilities and delays \cite{Zhang2019UnknownLosses,Hu2022DistributedFusion,Lv2023AdaptiveCIF,Tian2025RobustNonlinear}; these methods primarily address measurement availability rather than the authenticity of successfully received data. Robust Bayesian filters instead use latent reliability indicators, Gaussian mixtures, or Bernoulli--Gaussian models to identify contaminated measurements and adapt their statistics \cite{Yun2022RobustCKF,Fu2023RobustCKF,Jiang2025OutlierRobustMHE,Cui2025VariationalCKF}, typically using separately parameterized nominal and abnormal noise models. Additive covariance inflation has also appeared in modeling measurement ambiguity and externally injected noise \cite{Panakkal2015MeasurementModel,LeNy2014DifferentiallyPrivate}, but without jointly modeling communication availability and latent data authenticity.
	
	A complementary line of research concerns variational Bayesian (VB) adaptive filtering for unknown noise statistics. 
    Recursive VB methods have been developed to jointly infer system states and uncertain noise statistics, ranging from adaptive estimation of measurement-noise covariance \cite{4796261} to simultaneous adaptation to inaccurate process and measurement noise covariances \cite{Huang2018AdaptiveKF}, and have subsequently been extended to nonlinear state-space models \cite{Sarkka2013Nonlinear,Yu2019Nonlinear}. More recently, conjugate variational inference adaptive Kalman filter (CVIAKF) employs conjugate-computation  variational inference  (CVI) \cite{khan2017conjugate} for joint nonlinear state and noise estimation while retaining analytical inverse-Wishart covariance updates \cite{10818708}. These formulations, however, do not address the nonconjugacy introduced when the same unknown baseline covariance is shared by clean and additively corrupted measurement components.

To overcome these limitations, we propose a conjugate variational inference adaptive Kalman filter with dual reliability (CVIAKF-DR) built on structured CVI. Our prior work \cite{SWL:26} established a variational Bayesian framework for linear systems with intermittently corrupted observations, jointly modeling packet reception, measurement authenticity, and unknown noise statistics. This work extends the framework to nonlinear systems for joint state–noise inference and addresses nonconjugacy from nonlinear estimation and the shared additive covariance structure. Two main contributions are as follows. First, we extend the dual-reliability framework to nonlinear systems. Packet reception and latent measurement authenticity represent communication availability and data validity. Clean and corrupted measurements share a baseline covariance through \(\boldsymbol{R}_k\) and \(\boldsymbol{R}_k+\boldsymbol{E}_k\), distinguishing intrinsic measurement uncertainty from corruption-induced uncertainty and supporting joint nonlinear-state inference. Second, we devise a structured CVI algorithm for joint estimation of the state, prediction covariance, baseline measurement covariance, packet-reception probability, and measurement-authenticity probability. The prediction covariance and reliability probabilities retain closed-form conjugate updates, while CVI handles nonconjugate terms of the nonlinear state and baseline covariance. Posterior authenticity probabilities adaptively combine expected measurement precisions in state updates, weighting data by inferred reliability.
	
	The remainder of this paper is organized as follows. Section II presents the problem formulation and reviews CVIAKF. Section III derives CVIAKF-DR, with emphasis on the nonconjugate covariance update. Section IV reports numerical results, and Section V concludes the paper.
	
	\section{Problem Setting-up and Preliminaries}\label{sec:problem}
	\subsection{Problem Setting-up}
	
	Consider a time series of length $T$ for a nonlinear dynamic system. We consider a centralized network of $L$ distributed sensors tracking a single target to model large-scale distributed estimation with network packet dropouts. At time $k$, $\boldsymbol{x}_k\in\mathbb R^{d_x}$ denotes the system state and $\boldsymbol{y}_{i,k}\in\mathbb R^{d_y}$ is the measurement from sensor $i$. The state and observation models read
	\begin{align}\label{eq:nonlinear-state}
		\boldsymbol{x}_k=&\boldsymbol{f}_{k-1}(\boldsymbol{x}_{k-1})+\boldsymbol{w}_{k-1},\\\label{eq:received-model}
		\boldsymbol{y}_{i,k}=&\gamma_{i,k}\bigl[\boldsymbol{h}_k(\boldsymbol{x}_k)+\boldsymbol{v}_{i,k}
		+(1-z_{i,k})\boldsymbol{\epsilon}_{i,k}\bigr],
	\end{align}
	for $i=1,\dots,L$. Here $\boldsymbol{f}_{k-1}(\cdot),\boldsymbol{h}_k(\cdot)$ are known nonlinear mappings. The noises $\boldsymbol{w}_{k-1}\sim\N(0,\boldsymbol{Q}_{k-1})$, $\boldsymbol{v}_{i,k}\sim\N(0,\boldsymbol{R}_k)$, and corruption noise $\boldsymbol{\epsilon}_{i,k}\sim\N(0,\boldsymbol{E}_k)$ are mutually independent. $\boldsymbol{Q}_k$, $\boldsymbol{R}_k\succ0$ are unknown, while $\boldsymbol{E}_k\succeq0$ is fixed and known.
	
	The binary observable $\gamma_{i,k}\in\{0,1\}$ models packet reception: $\gamma_{i,k}=1$ means successful packet arrival with $\Pr(\gamma_{i,k}=1)=\rho_k$. Thus, $1-\rho_k$ is the packet-loss probability, i.e. $\gamma_{i,k}|\rho_k\sim\Bern(\rho_k)$, where $\Bern(p)$ is a Bernoulli distribution with parameter $p$.
	
	Even received packets may be corrupted; we introduce the latent binary indicator $z_{i,k}\in\{0,1\}$, where $z_{i,k}=1$ corresponds to clean measurements and $z_{i,k}=0$ indicates corruption with extra noise $\boldsymbol{\epsilon}_{i,k}$. Thus, $1-\beta_k$ is the corruption probability, i.e. $z_{i,k}\mid\beta_k,\gamma_{i,k}=1\sim\Bern(\beta_k)$.
	
	The measurement likelihood for received packets is
	\begin{align}
		&p(\boldsymbol{y}_{i,k}|\boldsymbol{x}_k,\boldsymbol{R}_k,z_{i,k},\gamma_{i,k}=1)\notag\\
		=&\N\!\left(\boldsymbol{y}_{i,k};\boldsymbol{h}_k(\boldsymbol{x}_k),
		\boldsymbol{R}_k+(1-z_{i,k})\boldsymbol{E}_k\right).
		\label{eq:mixture-model}
	\end{align}
	Thus $z_{i,k}$ selects either covariance $\boldsymbol{R}_k$ (clean) or $\boldsymbol{R}_k+\boldsymbol{E}_k$ (corrupted). When $\gamma_{i,k}=0$, no measurement likelihood is incorporated, and missing packets are not substituted with artificial zero observations.
	
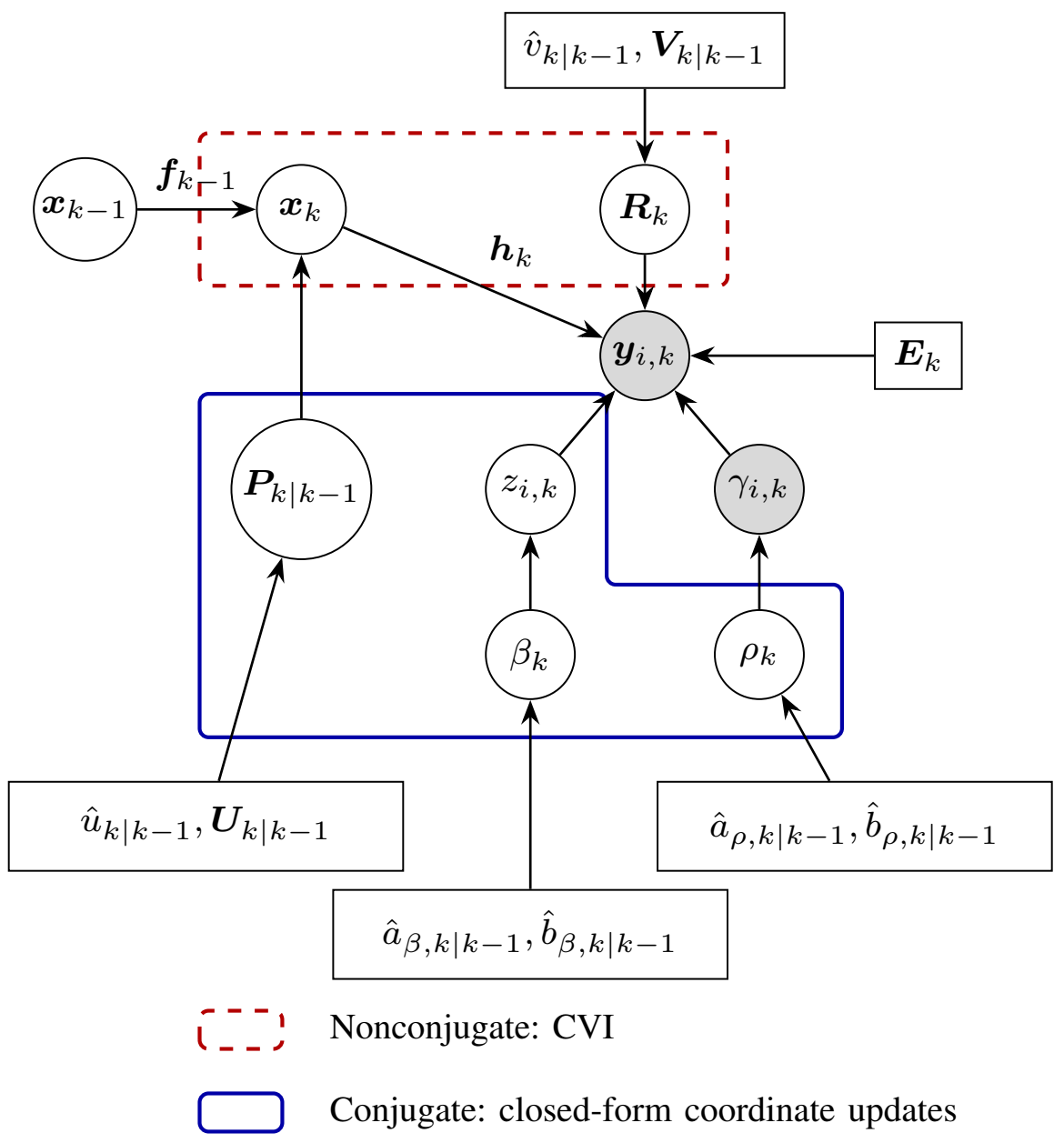
\begin{figure}[htbp]
	\centering
	\resizebox{0.88\columnwidth}{!}{%
		\begin{tikzpicture}[
			font=\footnotesize,
			>=Stealth,
			latent/.style={
				circle,
				draw,
				minimum size=7mm,
				inner sep=1pt
			},
			observed/.style={
				latent,
				fill=black!15
			},
			known/.style={
				rectangle,
				draw,
				inner sep=4pt
			},
			prior/.style={
				rectangle,
				draw,
				minimum width=31mm,
				minimum height=7mm,
				align=center,
				inner sep=3pt
			},
			link/.style={
				->,
				line width=.55pt
			},
			conjugate/.style={
				draw=blue!65!black,
				solid,
				line width=.85pt,
				rounded corners=2pt
			},
			nonconjugate/.style={
				draw=red!70!black,
				dashed,
				line width=.85pt,
				rounded corners=2pt
			}
		]
			
			\draw[nonconjugate]
			(-2.60,1.90) rectangle (1.55,3.10);

			\draw[conjugate]
			(-2.60,1.05)
			-- (0.60,1.05)
			-- (0.60,-0.45)
			-- (2.45,-0.45)
			-- (2.45,-1.65)
			-- (-2.60,-1.65)
			-- cycle;

			\node[latent] (xprev) at (-3.50,2.50)
			{$\boldsymbol{x}_{k-1}$};
			
			\node[latent] (x) at (-1.80,2.50)
			{$\boldsymbol{x}_k$};
			
			\node[latent] (R) at (0.90,2.50)
			{$\boldsymbol{R}_k$};
			
			\node[known] (Rprior) at (0.90,3.75)
			{$\hat{v}_{k|k-1},
			  \boldsymbol{V}_{k|k-1}$};

			\node[observed] (y) at (0.90,1.35)
			{$\boldsymbol{y}_{i,k}$};
			
			\node[known] (E) at (3.05,1.35)
			{$\boldsymbol{E}_k$};

			\node[latent,minimum size=11mm] (P) at (-1.80,0.30)
			{$\boldsymbol{P}_{k|k-1}$};
			
			\node[latent] (z) at (0,0.30)
			{$z_{i,k}$};
			
			\node[observed] (gamma) at (1.80,0.30)
			{$\gamma_{i,k}$};

			\node[latent] (beta) at (0,-1.00)
			{$\beta_k$};
			
			\node[latent] (rho) at (1.80,-1.00)
			{$\rho_k$};

			%
			\node[prior] (Pprior) at (-2.55,-2.35)
			{$\hat{u}_{k|k-1},
			  \boldsymbol{U}_{k|k-1}$};
			
			\node[prior] (rhoprior) at (2.55,-2.35)
			{$\hat{a}_{\rho,k|k-1},
			  \hat{b}_{\rho,k|k-1}$};
			
			\node[prior] (betaprior) at (0,-3.20)
			{$\hat{a}_{\beta,k|k-1},
			  \hat{b}_{\beta,k|k-1}$};

			\draw[link]
			(xprev) --
			node[above] {$\boldsymbol{f}_{k-1}$}
			(x);
			
			\draw[link]
			(Pprior) -- (P);
			
			\draw[link]
			(P) -- (x);
			
			\draw[link]
			(x) --
			node[above right] {$\boldsymbol{h}_k$}
			(y);
			
			\draw[link]
			(Rprior) -- (R);
			
			\draw[link]
			(R) -- (y);
			
			\draw[link]
			(E) -- (y);
			
			\draw[link]
			(betaprior) -- (beta);
			
			\draw[link]
			(beta) -- (z);
			
			\draw[link]
			(z) -- (y);
			
			\draw[link]
			(rhoprior) -- (rho);
			
			\draw[link]
			(rho) -- (gamma);
			
			\draw[link]
			(gamma) -- (y);

			\draw[nonconjugate]
			(-2.60,-4.10) rectangle (-1.95,-3.80);
			
			\node[anchor=west] at (-1.70,-3.95)
			{Nonconjugate: CVI};
			
			\draw[conjugate]
			(-2.60,-4.75) rectangle (-1.95,-4.45);
			
			\node[anchor=west] at (-1.70,-4.60)
			{Conjugate: closed-form coordinate updates};
			
		\end{tikzpicture}%
	}
	
	\caption{Generative model of the NLF problem
		\eqref{eq:nonlinear-state}--\eqref{eq:received-model}
		with packet dropouts and corrupted noises.
		Shaded nodes: observable variables; unshaded circles:
		latent parameters to be estimated; rectangles:
		hyper-parameters in priors.}
	\label{fig:generative-model}
\end{figure}

	Let
	\begin{equation}
		\mathcal I_k=\{i:\gamma_{i,k}=1\},\qquad M_k=|\mathcal I_k|.
		\label{eq:received-set}
	\end{equation}
	All $\gamma_{i,k}$ are observed and used to infer $\rho_k$, while only the $M_k$ received measurements inform $\boldsymbol{x}_k$, $\boldsymbol{R}_k$, $\beta_k$, and latent indicators $\{z_{i,k}\}_{i\in\mathcal I_k}$. The generative
	model and its known/unknown parameters are displayed in Fig.~\ref{fig:generative-model}.
	
	\subsection{Conjugate variational inference (CVI)}\label{sec-CVI}
	
	The CVI \cite{khan2017conjugate}  is proposed to resolve the computationally challenging in variational inference (VI) in models that contain both conjugate
	and non-conjugate terms. It uses a stochastic mirror descent method in the expectation parameter space,
	and then expressing each gradient step as a VI in a conjugate model. In this subsection we recall the CVI with the original notations in \cite{khan2017conjugate}.
	
	Given a probabilistic graphical model $p(\by, \bz)$ with $\by$ being the observed variables and $\bz$ being latent variables to be estimated, our goal in VI is to estimate a posterior distribution $p(\bz|\by)$ with a fixed-form variational approximation $q(\bz;\bl)$, by maximizing a lower bound to the marginal likelihood:
	\begin{align}\label{eqn-2.1}
		q(\bz;\bl^*):=\max_{\lambda\in\Omega}\cL(\bl):=\E_q[\log p(\by,\bz)-\log q(\bz;\bl)],
	\end{align}
	where $\bl$ is the natural parameter, $\Omega$ is the set of valid natural parameters. The CVI relies on the following two assumptions:
	\begin{assumption}[minimality]\label{as-1}
		The variational distribution $q(\mathbf{z};\boldsymbol{\lambda})$ is a ``minimal'' exponential-family distribution:
		\begin{equation}
			q(\mathbf{z};\boldsymbol{\lambda}) = h(\mathbf{z}) \exp\left\{\langle \boldsymbol{\phi}(\mathbf{z}), \boldsymbol{\lambda} \rangle - A(\boldsymbol{\lambda})\right\}.
			\label{eq:exp_family}
		\end{equation}
	\end{assumption}
	The minimal representation implies the existence of a one-to-one mapping between the expectation parameter $\boldsymbol{\mu} := \mathbb{E}_q[\boldsymbol{\phi}(\mathbf{z})]$ and the natural parameter $\boldsymbol{\lambda}$. 
	
	\begin{assumption}[conjugacy]\label{as-2}
		The joint distribution 
		$p(\mathbf{y},\mathbf{z}) \propto \tilde{p}_{nc}(\mathbf{y},\mathbf{z})\tilde{p}_c(\mathbf{y},\mathbf{z})$, where $\tilde{p}_c$ is the conjugate part of the mode, taking the same form as $q$ with respect to $\mathbf{z}$, i.e.,
		\begin{equation}
			\tilde{p}_c(\mathbf{y},\mathbf{z}) \propto h(\mathbf{z}) \exp\left\{\langle \boldsymbol{\phi}(\mathbf{z}), \boldsymbol{\eta} \rangle\right\},
			\label{eq:conjugate_part}
		\end{equation}
		where $\boldsymbol{\eta}$ is a known parameter vector; while $\tilde{p}_{nc}$ is the non-conjugate terms. These terms can be unnormalized with respect to $\mathbf{z}$.
	\end{assumption}
	
	
	\begin{theorem}[Theorem 1, \cite{khan2017conjugate}]\label{thm-1}
		Under Assumption \ref{as-1} and \ref{as-2}, the Bayesian inference can be written in the conjugate model:
		\begin{equation}
			q(\mathbf{z};\boldsymbol{\lambda}_{t}) \propto e^{\langle \boldsymbol{\phi}(\mathbf{z}), \widetilde{\boldsymbol{\lambda}}_t \rangle} \tilde{p}_c(\mathbf{y},\mathbf{z}),
			\label{eq:conj-model-11}
		\end{equation}
		whose natural parameter can be obtained by conjugate computation: $\boldsymbol{\lambda}_{t} = \widetilde{\boldsymbol{\lambda}}_t + \boldsymbol{\eta}$, where $\widetilde{\boldsymbol{\lambda}}_t$ is the natural parameters of the exponential-family approximation to $\tilde{p}_{nc}$, which can itself be obtained recursively by using stochastic gradients of $\tilde{p}_{nc}$:
		\begin{equation}
			\widetilde{\boldsymbol{\lambda}}_{t+1} = (1-\beta_t)\widetilde{\boldsymbol{\lambda}}_{t} + \beta_t \left. \nabla_{\boldsymbol{\mu}} \mathbb{E}_q\left[\log \tilde{p}_{nc}\right] \right|_{\boldsymbol{\mu}=\boldsymbol{\mu}_t},
			\label{eq:recur-lambda-12}
		\end{equation}
		with $\tilde \bl_0 = 0$ and $\bl_0=\boldsymbol{\eta}$.
	\end{theorem}

	\subsection{Joint estimation in CVIAKF}\label{sec-2.3}
	
	Following the framework of recursive Bayesian filtering, CVIAKF encompasses three essential phases: initialization, prediction, and update. The initialization phase involves modeling the prior PDF of latent variables. These three phases are solved recursively in cycle.
	\begin{enumerate}
		\item \textit{Initialization}: Assume that the joint conditional distribution of $\boldsymbol{\Theta}_{k-1}$ given measurements $\boldsymbol{y}_{1:k-1}$ is approximated as a product of distributions as follows:
		\begin{equation}
			\begin{aligned}
				& p(\boldsymbol{\Theta}_{k-1}|\boldsymbol{y}_{1:k-1}) \\
				=& \mathcal{N}\big(\boldsymbol{x}_{k-1}|\hat{\boldsymbol{x}}_{k-1|k-1}, \bP_{k-1|k-1}\big) \\
				&\times\IW\big(\bP_{k-1|k-1}|\hat{u}_{k-1|k-1}, \boldsymbol{U}_{k-1|k-1}\big) \\
				&\times\IW\big(\boldsymbol{R}_{k-1}|\hat{v}_{k-1|k-1}, \boldsymbol{V}_{k-1|k-1}\big).
			\end{aligned}
			\label{eq:init-14}
		\end{equation}
		\item \textit{Prediction}: Given the previous distibution in 1), $p(\boldsymbol{\Theta}_{k-1}|\boldsymbol{y}_{1:k-1})$, the predicted distribution can be factorized as
		\begin{equation}
			\begin{aligned}
				&p(\boldsymbol{\Theta}_k|\boldsymbol{y}_{1:k-1}) \\
				=& \cN\big(\boldsymbol{x}_k|\hat{\boldsymbol{x}}_{k|k-1}, \boldsymbol{P}_{k|k-1}\big) \IW\big(\boldsymbol{P}_{k|k-1}|\hat{u}_{k|k-1}, \boldsymbol{U}_{k|k-1}\big) \\
				&\times \IW\big(\boldsymbol{R}_k|\hat{v}_{k|k-1}, \boldsymbol{V}_{k|k-1}\big)
			\end{aligned}
			\label{eq:pred-new-16}
		\end{equation}
		with the parameters of the predicted distribution given by \cite{Huang2018AdaptiveKF}
		\begin{equation}
            \begin{aligned}
			\hat{\boldsymbol{x}}_{k|k-1} 
			= &\mathbb{E}\big[\boldsymbol{f}_k(\boldsymbol{x}_{k-1})\big],\\
			\boldsymbol{P}_{k|k-1} 
			= &\mathbb{E}\Big[\big(\boldsymbol{f}_k(\boldsymbol{x}_{k-1}) - \hat{\boldsymbol{x}}_{k|k-1}\big)(\cdot)^\top\Big] + \hat{\boldsymbol{Q}}_k,\\
			\hat{u}_{k|k-1} 
			=& d_x+ \tau_P + 1,\quad \boldsymbol{U}_{k|k-1} = \tau_P \boldsymbol{P}_{k|k-1},\\
			\hat{v}_{k|k-1} =& \varrho_R \big(\hat{v}_{k-1|k-1} - d_y- 1\big) + d_y + 1, \\
			\boldsymbol{V}_{k|k-1}=& \varrho_R \boldsymbol{V}_{k-1|k-1}.
			\label{eq:iwr-param-19}
		\end{aligned}
        \end{equation}
		where $\hat{\boldsymbol{Q}}_k$ is the nominal process noise covariance, $\tau_P > 0$ is a tuning parameter, and $\varrho_R \in (0,1]$ is the forgetting factor of $\boldsymbol{R}_k$. The initial $\boldsymbol{R}_0$ is assumed to be IW distribution, i.e., $p(\boldsymbol{R}_0) = \IW(\boldsymbol{R}_0;\hat{v}_{0|0},\boldsymbol{U}_{0|0})$. The expectation of $\boldsymbol{R}_0$ satisfies $\hat{\boldsymbol{R}}_0 = \boldsymbol{V}_{0|0}/(\hat{v}_{0|0}-d_y-1)$ with $\hat{v}_{0|0} = \tau_R + d_y + 1$ and $\boldsymbol{V}_{0|0} = \tau_R \hat{\boldsymbol{R}}_0$.
		
		\item[3)] \textit{Update}: By the assumption that the likelihood function is a Gaussian distribution, the predicted PDF $p(\boldsymbol{\Theta}_k|\boldsymbol{y}_{1:k-1})$ is updated with measurement $\boldsymbol{y}_k$ by
		\begin{equation}
			\begin{aligned}
				&p(\boldsymbol{\Theta}_k|\boldsymbol{y}_{1:k}) \\
				\propto& \mathcal{N}\big(\boldsymbol{y}_k|\boldsymbol{h}_k(\boldsymbol{x}_k), \boldsymbol{R}_k\big)\mathcal{N}\big(\boldsymbol{x}_k|\hat{\boldsymbol{x}}_{k|k-1}, \boldsymbol{P}_{k|k-1}\big) \\
				&\times \IW\big(\boldsymbol{P}_{k|k-1}|\hat{u}_{k|k-1}, \boldsymbol{U}_{k|k-1}\big) \\
				&\times\IW\big(\boldsymbol{R}_k|\hat{v}_{k|k-1}, \boldsymbol{V}_{k|k-1}\big).
			\end{aligned}
			\label{eq:update-20}
		\end{equation}
	\end{enumerate}

	The task of CVIAKF \cite{10818708} is  approximate the joint posterior distribution $p(\bTheta_k|\by_{1:k})$ of latent variables $\bTheta_k \triangleq \{\boldsymbol{x}_k, \bP_{k|k-1}, \boldsymbol{R}_k\}$  by an optimal mean-field variational distribution 
	\begin{equation}
		q(\bTheta_k;\bl_k) = q(\boldsymbol{x}_k;\bl_k^{\boldsymbol{x}})\, q(\bP_{k|k-1};\bl_k^{\bP})\, q(\boldsymbol{R}_k;\bl_k^{\boldsymbol{R}}),
		\label{eq:meanfield-5}
	\end{equation}
	minimizing Kullback--Leibler (KL) divergence, i.e.  ELBO $\cL(\bl_k)$ in \eqref{eqn-2.1}, 
	\begin{align*}
		&q(\bTheta_k;\bl_k^*)\\
		=& \mathop{\arg\max}_{\bl_k} \underbrace{\mathbb{E}_{q(\bTheta_k)}\big[\log p(\bTheta_k, \by_k)-\log q(\bTheta_k;\bl_k)\big]}_{\cL(\bl_k)},
	\end{align*}
	where $\bl_k=(\bl_k^{\boldsymbol{x}}, \bl_k^{\bP}, \bl_k^{\boldsymbol{R}})$ is the natural parameters. 
	By \eqref{eq:update-20}, both $\bP_{k|k-1}$ and $\bR_k$ are conjugate, while $\bx_k$ is nonconjugate. The posterior parameters are updated coordinately. Due to the page limit, we refer the readers to Theorem 2, \cite{10818708}.

	\section{conjugate variational inference adaptive Kalman filter with dual reliability (CVIAKF-DR)}\label{sec:inference}
	
	In our problem, we shall estimate the state, noises and the two reliability probabilities $\Xi_k:=\{\bx_k,\bP_{k|k-1},\bR_k,\rho_k,\beta_k,
	z_{\mathcal I_k,k}\}$ condition on the observation history and the package receipt indicators $\mathcal Y_k=\{\boldsymbol{y}_{i,k}:\,i\in\mathcal I_k\}$ and $\{\gamma_{i,k},\ i\in\mathcal I_k\}$. 

    As in CVIAKF, the initialization at time instant $k$ is given by the posterior distributions of the parameters at previous time instant $k-1$, i.e.
\begin{align*}
    &p(\Xi_{k-1}|\by_{1:k-1})\\
    =&\textup{R.H.S of }\eqref{eq:init-14}\times\Beta(\rho_{k-1};\hat{a}_{\rho,k-1|k-1},\hat{b}_{\rho,k-1|k-1})\\ \notag
    &\times\Beta(\beta_{k-1};\hat{a}_{\beta,k-1|k-1},\hat{b}_{\beta,k-1|k-1})\\
    &\times\Bern(z_{i,k-1};\hat{\omega}_{i,k-1}),
\end{align*}
and the prediction is as in \eqref{eq:pred-new-16} 
\begin{align}\notag\label{eqn-3.1}
    &p(\Xi_k|\by_{1:k-1})\\\notag
    =&\textup{R.H.S. of }\eqref{eq:pred-new-16}\times \Beta(\rho_{k};\hat{a}_{\rho,k|k-1},\hat{b}_{\rho,k|k-1})\\ \notag
    &\times\Beta(\beta_{k};\hat{a}_{\beta,k|k-1},\hat{b}_{\beta,k|k-1})\\
    &\times\Bern(z_{i,k};\hat{\omega}_{i,k}),
\end{align}
where the parameters of the predicted distribution given in \eqref{eq:iwr-param-19}:
\begin{align}
    \hat{a}_{\cdot,k|k-1}
    &=1+\varrho_\cdot(\hat{a}_{\cdot,k-1|k-1}-1),\notag\\
    \hat{b}_{\cdot,k|k-1}
    &=1+\varrho_\cdot(\hat{b}_{\cdot,k-1|k-1}-1),
    \label{eq:beta-forgetting-rho}
\end{align}
where the subscription $\cdot=\rho$ or $\beta$, $\varrho_\cdot\in(0,1]$ is the forgetting factor as $\varrho_R$ before. Thus, the update with measurement $\by_k$ is 
\begin{align*}
    p(\Xi_k|\by_{1:k)}\propto&\cN(\by_k|\bh_k(\bx_k),\bR_k)\times\textup{R.H.S. of }\eqref{eqn-3.1}.
\end{align*}
The mean-field variational distribution is as in \eqref{eq:meanfield-5}:
\begin{align}\notag
    q(\Xi_k;\bl_k)
    = &q(\boldsymbol{x}_k;\bl_k^{\boldsymbol{x}})\, q(\bP_{k|k-1};\bl_k^{\bP})\, q(\boldsymbol{R}_k;\bl_k^{\boldsymbol{R}}) \\
    &\times q(\rho_k,\bl_k^\rho)\,q(\beta_k,\bl_k^\beta)\,\prod_{i\in\cI_k}q(z_{i,k},\bl_{i,k}^z) \notag\\
    =&\N(\bx_k;\hat{\boldsymbol{x}}_{k|k},\boldsymbol{P}_{k|k})\,\IW(\bP_{k|k-1};\hat{u}_{k|k},\boldsymbol{U}_{k|k})\notag\\ 
    &\times\IW(\bR_k;\hat{v}_{k|k},\boldsymbol{V}_{k|k}) \Beta(\rho_k;\hat{a}_{\rho,k|k},\hat{b}_{\rho,k|k}) \notag\\
    &\times\Beta(\beta_k;\hat{a}_{\beta,k|k},\hat{b}_{\beta,k|k})\prod_{i\in\cI_k}\Bern(z_{i,k};\hat{\omega}_{i,k}),
    \label{eq:meanfield-all}
\end{align}  
where $\bl_k^\cdot$ and $\cdot_{k|k}$ are two sets of parameters. Either one is updated coordinately by maximizing its corresponding part of ELBO, due to its conjugacy/nonconjugacy. 
\begin{align}\label{eq:expanded-elbo}
		\mathcal L(\bl_k)={}&\E_q\log p(\boldsymbol{x}_k|\boldsymbol{P}_{k|k-1})
		+\E_q\log p(\boldsymbol{P}_{k|k-1})\notag\\
		&+\E_q\log p(\boldsymbol{R}_k)+\E_q\log p(\rho_k)+\E_q\log p(\beta_k)\notag\\
		&+\sum_{i=1}^{L}\E_q[\gamma_{i,k}\log\rho_k
		+(1-\gamma_{i,k})\log(1-\rho_k)]\notag\\
		&+\sum_{i\in\mathcal I_k}\E_q[z_{i,k}\log\beta_k
		+(1-z_{i,k})\log(1-\beta_k)]\notag\\\notag
		&+\sum_{i\in\mathcal I_k}\E_q\log\N(\boldsymbol{y}_{i,k};\boldsymbol{h}_k(\boldsymbol{x}_k),
		\boldsymbol{R}_k+(1-z_{i,k})\boldsymbol{E}_k)\\
		&-\E_q\log q(\Xi_k;\bl_k).
	\end{align}

	\subsection{Variational Updates}
	\label{sec:update-theorem}

In this subsection, we shall detail the variational update by maximizing the ELBO \eqref{eq:expanded-elbo} coordinate-wisely at time instant $k$.
	
\subsubsection{Inference for $\rho_k$}
	\label{sec:proof-rates}

Due to the conjugacy and the full knowledge of $\gamma_{i,k}$, we can easily derive the posterior of $\rho_k$, without any iteration as we shall do for other parameters:
	\begin{equation}
			\hat{a}_{\rho,k|k}=\hat{a}_{\rho,k|k-1}+M_k,\quad
			\hat{b}_{\rho,k|k}=\hat{b}_{\rho,k|k-1}+L-M_k.
			\label{eq:rho-update}
		\end{equation}
        
\subsubsection{Inference for $z_{i,k},\,i\in\cI_k$ and $\beta_k$}
	\label{sec:proof-correctness}

If $z_{i,k}$ is known as $\gamma_{i,k}$ in Section \ref{sec:proof-rates}, then $\beta_k$ can be updated easily as $\rho_k$ in \eqref{eq:rho-update}. So we need to inference $z_{i,k}$ first. According to \eqref{eq:expanded-elbo}, the corresponding part of the EBLO involving $z_{i,k}$, for each $i\in\cI_k$:
\begin{align}
 \mathcal{L}(\hat\omega_{i,k})
 =&\E_q[z_{i,k}\log\beta_k
+(1-z_{i,k})\log(1-\beta_k)] \notag\\
&+\E_q\log\N\bigl(
\boldsymbol{y}_{i,k};\boldsymbol{h}_k(\boldsymbol{x}_k),\boldsymbol{R}_k+(1-z_{i,k})\boldsymbol{E}_k)\notag\\
&-\E_q \log q(z_{i,k}|\hat\omega_{i,k})\notag \\       =&\hat\omega_{i,k}\Delta_{i,k}^{1,(r)}+(1-\hat\omega_{i,k})\Delta_{i,k}^{0,(r)}\notag\\
		&-\hat\omega_{i,k}\log\hat\omega_{i,k}-(1-\hat\omega_{i,k})\log(1-\hat\omega_{i,k})+\mathrm{const}_\omega,
 \label{eq:z-coordinate-density}
\end{align}
where $\mathrm{const}_\omega$ is a constant independent of $\omega$, and 
	\begin{align}
		\Delta_{i,k}^{1,(r)}
		&=\E_{q_\beta^{(r)}}\log\beta_k+\E_{q_x^{(r)}q_R^{(r)}}\log\N(\boldsymbol{y}_{i,k};\boldsymbol{h}_k(\boldsymbol{x}_k),\boldsymbol{R}_k),\notag\\
		\Delta_{i,k}^{0,(r)}
		&=\E_{q_\beta^{(r)}}\log(1-\beta_k)\notag\\
		&\quad+\E_{q_x^{(r)}q_R^{(r)}}\log\N(\boldsymbol{y}_{i,k};\boldsymbol{h}_k(\boldsymbol{x}_k),\boldsymbol{R}_k+\boldsymbol{E}_k),
		\label{eq:branch-scores}
	\end{align}
where the superscription $(r)$ is the iteration index. The initialization of the iteration is the prediction distribution at time instant $k$. 

By directly taking its gradient with respect to $\hat\omega_{i,k}$ and letting it be zero, one has
		\begin{equation}
			\hat{\omega}_{i,k}^{(r+1)}=
			\left\{1+\exp\left[-\left(\Delta_{i,k}^{1,(r)}-
				\Delta_{i,k}^{0,(r)}\right)\right]\right\}^{-1}.
			\label{eq:omega-update}
		\end{equation}

After we update $z_{i,k}$, similar as in Section \ref{sec:proof-rates}, the inference of $\beta_k$ follows immediately by the conjugacy:
\begin{align}
			\label{eq:beta-update}
	\hat{a}_{\beta,k|k}^{(r+1)}&=\hat{a}_{\beta,k|k-1}+\sum_{i\in\mathcal I_k}\hat{\omega}_{i,k}^{(r+1)},\notag\\
	\hat{b}_{\beta,k|k}^{(r+1)}&=\hat b_{\beta,k|k-1}+\sum_{i\in\mathcal I_k}\left[1-\hat{\omega}_{i,k}^{(r+1)}\right].
\end{align}
	
\subsubsection{Inference for $\boldsymbol{R}_k$}
	\label{sec:proof-R}

According to \eqref{eq:expanded-elbo}, the corresponding part involving $\boldsymbol{R}_{k}$ is
	\begin{align}\notag
		&\mathcal{L}(\bl_k^{\bR})\\ =&\log\IW(\boldsymbol{R}_k;\hat{v}_{k|k-1},\boldsymbol{V}_{k|k-1})\notag\\
		&\underbrace{\phantom{aaaaaaaaa}+\sum_{i\in\mathcal I_k}\hat{\omega}_{i,k}^{(r+1)}\E_{q_x^{(r)}}\log\N(\boldsymbol{y}_{i,k};\boldsymbol{h}_k(\boldsymbol{x}_k),\boldsymbol{R}_k)}_{\log\tilde p^{(r)}_{c,k}(\bR_k,\by_{i,k})}\notag\\\notag
		&+\underbrace{\sum_{i\in\mathcal I_k}(1-\hat{\omega}_{i, k}^{(r+1)})\E_{q_x^{(r)}}
		\log\N(\boldsymbol{y}_{i,k};\boldsymbol{h}_k(\boldsymbol{x}_k),\boldsymbol{R}_k+\boldsymbol{E}_k)}_{\log\tilde p^{(r)}_{nc,k}(\bR_k,\by_{i,k})}\\
        &-\E_q\log q^{(r)}(\bR_k;\bl_k^{\bR})+\mathrm{const}.
		\label{eq:r-exact-coordinate}
	\end{align}
	The third term on the right-hand side of \eqref{eq:r-exact-coordinate}, the corrupted factor $\boldsymbol{R}_k+\boldsymbol{E}_k$, prevents the conjugacy, i.e. the posterior of $\boldsymbol{R}_k$
	from being inverse-Wishart. Therefore, the CVI \cite{khan2017conjugate} is used to deal with this issue. 

Recall Assumption \ref{as-1}-\ref{as-2} in Section \ref{sec-CVI},  $q^{(r)}(\bR_k;\bl_k^{\bR})=\IW(\boldsymbol{R}_k;\hat{v}_{k|k}^{(r)},\boldsymbol{V}_{k|k}^{(r)})$ is a minimal exponential-family distribution, with the sufficient statistics and natural parameters 
\begin{equation}
    \phi^{\bR}=\begin{bmatrix}\log|\boldsymbol{R}_k|\\\boldsymbol{R}_k^{-1}\end{bmatrix},\quad	\boldsymbol{\lambda}_{k}^{\bR,(r)}=-\frac12\begin{bmatrix}\hat{v}_{k|k}^{(r)}+d_y+1\\\boldsymbol{V}_{k|k}^{(r)}
        \end{bmatrix},
		\label{eq:r-sufficient-statistic}
	\end{equation}
respectively, and its expectation parameters is $\boldsymbol{\mu}_{ k}^{\bR,(r)}=\E_{q_R^{(r)}}\phi^{\bR}$. The natural parameter of the conjugate part $\tilde p_{c,k}^{(r)}$ is explicitly given by
	\begin{align}\label{eq:r-conjugate-message}
		&\boldsymbol{\eta}_{ k}^{\bR,(r)}\\\notag
        =&-\frac12
		\begin{bmatrix}
			\hat{v}_{k|k-1}+d_y+1+\displaystyle\sum_{i\in\mathcal I_k}\hat{\omega}_{i,k}^{(r+1)}\\
			\boldsymbol{V}_{k|k-1}+\displaystyle\sum_{i\in\mathcal I_k}\hat{\omega}_{i,k}^{(r+1)}
				\E_{q_x^{(r)}}[\,(\boldsymbol{y}_{i,k}-\boldsymbol{h}_k(\boldsymbol{x}_k))(\cdot)]^T		
		\end{bmatrix}.
	\end{align}
By Theorem \ref{thm-1}, under Assumption \ref{as-1}-\ref{as-2}, the Bayesian inference written in the conjugate model is 
\[
    q^{(r)}\left(\bR_k;\bl_{k,T_{\bR}}^{\bR,(r)}\right)\propto e^{\left\langle \phi^{\bR},\tilde \bl^{\bR,(r)}_{k,T_{\bR}}\right\rangle} \tilde p^{(r)}_{c,k}(\bR_k,\by_{\cI_k,k}),
\]
where the natural parameter $\bl_{k,T_{\bR}}^{\bR,(r)}$ is obtained by iteration from $t=0,\cdots,T_{\bR}$ with the step size $\alpha_{\bR,t}$
    \begin{align}   
    \boldsymbol{\lambda}_{k,t+1}^{\bR,(r)} &= (1-\alpha_{\bR,t} )\boldsymbol{\lambda}_{k,t}^{\bR,(r)}+\alpha_{\bR,t}\tilde \bl_{k,t}^{{\bR},(r)}\label{eq:r-update} 
    \end{align}
with
\begin{align}\label{eqn-3.2}
   \tilde \bl_{k,t}^{{\bR},(r)}
		&=\boldsymbol{\eta}_{ k}^{\bR,(r)}+
		\left.{\nabla}_{\bmu}
		\E_{q_R^{(r)}}[\log\tilde{p}_{nc,k}^{(r)}]
		\right|_{\bmu=\bmu_{k,t}^{\bR,(r)}},
    \end{align}
starting by $\tilde\bl_{k,0}^{\bR,(r)}=0$, $\bl_{k,0}^{\bR,(r)}=\boldsymbol{\eta}_{ k}^{\bR,(r)}$.
     
Since the gradient with respect to $\bmu$ in \eqref{eqn-3.2}  is always intractable, so we approximate it by Monte Carlo (MC):
\begin{align}\label{eq:r-expectation-gradient}
    \nabla_{\boldsymbol{\mu}}\E_{q_R^{(r)}}\left[\log\widetilde p_{{\rm nc},k}^{(r)}\right]=&\mathcal F_R^{-1}		\nabla_{\boldsymbol{\lambda}}\E_{q_R^{(r)}}\left[\log\tilde p_{nc,k}^{(r)}\right]\\\notag
    =&\mathcal F_R^{-1}\E_{q_R^{(r)}}\left[(\phi^{\bR}-\boldsymbol{\mu}_{ k,t}^{\bR,(r)})\log\tilde{p}_{nc,k}^{(r)}\right],
\end{align}
where $\mathcal F_R=\operatorname{Cov}_{q_R^{(r)}}(\phi^{\bR},\phi^{\bR})$ is the Fisher matrix. Consequently, the parameters $\hat{v}^{(r+1)}_{k|k}, \boldsymbol{V}^{(r+1)}_{k|k}$ of $\boldsymbol{R}_k$ are updated by $\bl_{k}^{\bR,(r+1)}=\bl_{k, T_{\bR}}^{\bR,(r)}$ and \eqref{eq:r-sufficient-statistic}
.

\subsubsection{Inference for $\boldsymbol{x}_k$}
	\label{sec:proof-state-P}

The inferences for these two parameters are similar to those in Theorem 2, \cite{10818708}. We only state the results directly for the completeness. The complete proof can be found in Appendix C, \cite{10818708}.
	
Due to the non-conjugacy caused by the nonlinearity of $h_k(\bx_k)$, the CVI has been used for $\bx_k$. The iteration index is $t'=0,\dots,T_{\bx}$. The state estimation update is:
		\begin{align}\notag
			\left(\boldsymbol{P}_{k|k,t'+1}^{(r)}\right)^{-1}
			=&(1-\alpha_{\bx,t'})\left(\boldsymbol{P}_{k|k,t'}^{(r)}\right)^{-1}\\
		&+\alpha_{\bx,t'}\left(\mathbb{E}[\boldsymbol{P}_{k|k-1}^{-1}]
        +\nabla_{\boldsymbol{P}_{k|k,t'}^{(r)}}\mathcal{D}(\bmu^{\bx,(r)}_{k,t'})\right)\notag\\
			&+\frac{\alpha_{x,t'}^2}{2}
			\widehat{\boldsymbol{G}}_{x,t'}^{(r)}\boldsymbol{P}_{k|k,t'}^{(r)}\widehat{\boldsymbol{G}}_{x,t'}^{(r)},
			\label{eq:x-cov-update}\\\notag
		\hat{\boldsymbol{x}}_{k|k,t'+1}^{(r)}
			=&\hat{\boldsymbol{x}}_{k|k,t'}^{(r)}
			-\frac{\alpha_{\bx,t'}}2\boldsymbol{P}_{k|k,t'+1}^{(r)}
			\nabla_{\hat{\boldsymbol{x}}_{k|k,t'}^{(r)}}\mathcal{D}(\bmu^{\bx,(r)}_{k,t'})\notag\\
			&+\alpha_{\bx,t'}\boldsymbol{P}_{k|k,t'+1}^{(r)}
			\mathbb{E}[\boldsymbol{P}_{k|k-1}^{-1}]
			\bigl(\hat{\boldsymbol{x}}_{k|k-1}-\hat{\boldsymbol{x}}_{k|k}^{(r)}\bigr),
			\label{eq:x-mean-update}
		\end{align}
		with the positive-definite compensation for the update $(\boldsymbol{P}_{k|k}^{(r)})^{-1}$:
		\begin{align*}
			\widehat{\boldsymbol{G}}_{x,t'}^{(r)}
			=&\sym\!\left((\boldsymbol{P}_{k|k,t'}^{(r)})^{-1}
			-\mathbb{E}[\boldsymbol{P}_{k|k-1}^{-1}]
			-\nabla_{\boldsymbol{P}_{k|k,t'}^{(r)}}\mathcal{D}(\bmu^{\bx,(r)}_{k,t'})\right),
		\end{align*}
		where 
		$\alpha_{\bx,t'}\in(0,1]$ is the step size,  $\sym(B)=(B+B^T)/2$, and the expectation
		$\mathbb{E}[\boldsymbol{P}_{k|k-1}^{-1}] =(\hat{u}_{k|k}^{(r)}-d_x-1)(\boldsymbol{U}_{k|k}^{(r)})^{-1}$ with $\hat{u}_{k|k}^{(0)}=\hat{u}_{k|k-1},\, \boldsymbol{U}_{k|k}^{(0)} = \boldsymbol{U}_{k|k-1} $.
		
The gradients in \eqref{eq:x-cov-update}-\eqref{eq:x-mean-update} are estimated by MC using the reparameterization trick \cite{SMMA:20} to reduce the high variance, by letting
		$\boldsymbol{x}_{s,k|k,t'}^{(r)}=\hat{\boldsymbol{x}}_{k|k,t'}^{(r)}+\boldsymbol{L}_{x,k|k,t'}^{(r)}\boldsymbol{\varepsilon}_s$ with
		$\boldsymbol{L}_{x,k|k,t'}^{(r)}(\boldsymbol{L}_{x,k|k,t'}^{(r)})^T=\boldsymbol{P}_{k|k,t'}^{(r)}$ and
		$\boldsymbol{\varepsilon}_s\sim\N(0,I)$, for $s=1,\ldots,S_x$, where $S_x$ is the
		number of samples.  Then
		\begin{align}\notag
			\nabla_{\hat{\boldsymbol{x}}_{k|k,t'}^{(r)}}\mathcal{D}(\boldsymbol{\mu}_{x,k,t'}^{(r)})  
			&\approx -\frac{2}{S_x}\sum_{s=1}^{S_x}
			\sum_{i\in\mathcal{I}_k}
			\boldsymbol{H}_k(\boldsymbol{x}_{s,k|k,t'}^{(r)})^{\mathrm{T}}\boldsymbol{W}_{i,k}^{(r+1)}\\
            &\phantom{aaaaaa}\times\left[\boldsymbol{y}_{i,k}-\boldsymbol{h}_k(\boldsymbol{x}_{s,k|k,t'}^{(r)})\right],
			\label{eq:x-grad-mean}
     \end{align}
    \begin{align}\label{eq:x-grad-cov}
			\notag&\nabla_{\boldsymbol{P}_{k|k,t'}^{(r)}}\mathcal{D}(\bmu^{\bx,(r)}_{k,t'})\\
			&\approx -\frac{1}{S_x}\sum_{s=1}^{S_x}
			\sum_{i\in\mathcal{I}_k}
			\boldsymbol{H}_k(\boldsymbol{x}_{s,k|k,t'}^{(r)})^{\mathrm{T}}\boldsymbol{W}_{i,k}^{(r+1)}\\\notag
			&\phantom{aaaaaaaaaaaaa}\times\bigl(\boldsymbol{y}_{i,k}-\boldsymbol{h}_k(\boldsymbol{x}_{s,k|k,t'}^{(r)})\bigr)
			\boldsymbol{\varepsilon}_s^{\mathrm{T}}(\boldsymbol{L}_{x,k|k,t'}^{(r)})^{-1},
		\end{align}
		where $\boldsymbol{H}_k(\boldsymbol{x})=\partial \boldsymbol{h}_k(\boldsymbol{x})/\partial \boldsymbol{x}$ and 
        \begin{align}\label{eqn-3.3}
           & \boldsymbol{W}_{i,k}^{(r+1)}\\\notag
		=&\hat{\omega}_{i, k}^{(r+1)}\E_{q_R^{(r+1)}}[\boldsymbol{R}_k^{-1}]
		+(1-\hat{\omega}_{i,k}^{(r+1)})
		\E_{q_R^{(r+1)}}[(\boldsymbol{R}_k+\boldsymbol{E}_k)^{-1}].
        \end{align}

We emphasize that this \(\boldsymbol{W}_{i,k}^{(r+1)}\) is the only difference for the \(\boldsymbol{x}_k\) inference between ours and that in Theorem 2, \cite{10818708}, which uses \(\mathbb{E}_{q_R^{(r+1)}}[\boldsymbol{R}_k^{-1}]\).

Consequently, the state parameters are updated by \((\boldsymbol{P}_{k|k}^{(r+1)})^{-1} = (\boldsymbol{P}_{k|k,T_{\boldsymbol{x}}}^{(r)})^{-1}\) and \(\hat{\boldsymbol{x}}_{k|k}^{(r+1)} = \hat{\boldsymbol{x}}_{k|k,T_{\boldsymbol{x}}}^{(r)}\), where \(T_{\boldsymbol{x}}\) represents the termination step.

    \subsubsection{Inference  for $\boldsymbol{P}_{k|k-1}$}\label{sec-sec:proof-P}
	
Due to the conjugacy, the update is derived through VI directly:	
		\begin{align}
			\hat{u}_{k|k}^{(r+1)}&=\hat{u}_{k|k-1}+1,\notag\\
			\boldsymbol{U}_{k|k}^{(r+1)}&=\boldsymbol{U}_{k|k-1}+\boldsymbol{P}_{k|k}^{(r+1)}\notag\\
			&\quad+(\hat{\boldsymbol{x}}_{k|k}^{(r+1)}-\hat{\boldsymbol{x}}_{k|k-1})(\hat{\boldsymbol{x}}_{k|k}^{(r+1)}-\hat{\boldsymbol{x}}_{k|k-1})^T.
			\label{eq:p-update}
		\end{align}

\begin{theorem}[CVIAKF-DR]
Considering the NLF problem \ref{eq:nonlinear-state} with $L$ observations in \eqref{eq:received-model}, where the packet receipt satisfies $\gamma_{i,k} \sim \Bern(\rho_k)$ and the authenticity of received measurement obeys $z_{i,k}|\beta_k,\gamma_{i,k}=1 \sim \Bern(\beta_k)$. The packet dropout probability $1-\rho_k$ and corrupted measurement probability $1-\beta_k$ are unknown.

Under Assumptions \ref{as-1}-\ref{as-2}, the optimal parameters approximating the posterior distribution of $\Xi_k \triangleq \{\boldsymbol{x}_k, \boldsymbol{P}_{k|k-1}, \boldsymbol{R}_k, \rho_k, \beta_k, z_{\mathcal{I}_k,k}\}$ are updated iteratively in the following sequence:
\begin{enumerate}[label=\arabic*)]
    \item Update the parameter of $\rho_k$ via  \eqref{eq:rho-update};
    \item Update the parameter of $z_{\mathcal{I}_k,k}$ via \eqref{eq:omega-update};
    \item Update the parameter of $\beta_k$ via \eqref{eq:beta-update};
    \item Update the natural parameter of $\boldsymbol{R}_k$ via  \eqref{eq:r-update}-\eqref{eqn-3.2};
    \item Update the parameter of $\boldsymbol{x}_k$ via \eqref{eq:x-cov-update}-\eqref{eq:x-mean-update};
    \item Update the parameter of $\boldsymbol{P}_{k|k-1}$ via \eqref{eq:p-update}.
\end{enumerate}
\end{theorem}
	
	\subsection{Algorithm}

    Algorithm \ref{alg:complete-cvi} summarizes the complete recursion of the proposed CVIAKF-DR at time instant $k$.
The algorithm adopts a double-loop iterative variational inference framework: the outer loop implements joint variational updates for all latent variables, while two nested inner loops perform block-wise CVI updates for the measurement noise covariance $\boldsymbol{R}_k$ and state $\boldsymbol{x}_k$ respectively. The outer iteration terminates when the convergence criterion 
\begin{align}\label{eq:stopping-rule}
		&\Delta_{\rm all}\\\notag
        :=&\max\{\delta(\hat{\boldsymbol{x}}_{k|k}),\delta(\boldsymbol{P}_{k|k}),\delta(\hat{u}_{k|k}),\delta(\boldsymbol{U}_{k|k}),
		\delta(\hat{v}_{k|k}),\delta(\boldsymbol{V}_{k|k}),\\\notag
		&\phantom{aaaaa}\delta(\hat{a}_{\beta,k|k}),\delta(\hat{b}_{\beta,k|k}),
		\|\boldsymbol{\hat{\omega}}_{k|k}^{(r+1)}-\boldsymbol{\hat{\omega}}_{k|k}^{(r)}\|_\infty\},
\end{align}
with $\delta(M) := \frac{\left\|M^{(r+1)} - M^{(r)}\right\|}{1+\left\|M^{(r)}\right\|}$, is met or the maximum outer iteration count $I_{\text{max}}$ is exhausted. For each inner CVI block, the iteration stops either upon satisfying 
\begin{equation}
\delta_{b,\ell} = \frac{\left\|\boldsymbol{\lambda}_{k,\ell+1}^{b,(r)} - \boldsymbol{\lambda}_{k,\ell}^{b,(r)}\right\|_2}{1+\left\|\boldsymbol{\lambda}_{k,\ell}^{b,(r)}\right\|_2}.
\label{eq:delta_block}
\end{equation}
or reaching the prescribed maximum inner updates $T_{\bR}$ and $T_{\bx}$. 

Note that the posterior of the packet survival probability is directly computed via a one-pass exact update given by \eqref{eq:rho-update} and is not involved in the iterative loops.

\begin{algorithm}[h!]
	\caption{CVIAKF-DR}
	\label{alg:complete-cvi}
	\KwIn{
		$\widehat{\boldsymbol{x}}_{k-1|k-1},
		\boldsymbol{P}_{k-1|k-1}$;
		$\gamma_{1:L,k}$;
		$\{\boldsymbol{y}_{i,k}\}_{i\in\mathcal I_k}$;
		and fixed $\boldsymbol{E}_k$.
	}
	\KwOut{
		$\widehat{\boldsymbol{x}}_{k|k},
		\boldsymbol{P}_{k|k},
		\boldsymbol{P}_{k|k-1},
		\boldsymbol{R}_{k|k},
		\widehat{\rho}_{k|k},
		\widehat{\beta}_{k|k}$.
	}
	\BlankLine

	Construct the prior distributions via \eqref{eqn-3.1}\;
	Initialize $q_x,q_P,q_R,q_\beta$ from their priors\;
	Update $q_\rho$ once using \eqref{eq:rho-update}\;

	\For{$r=0,1,\ldots,I_{\max}-1$}{
		Update each unfixed $\hat{\omega}_{i,k}^{(r+1)}$ by
		\eqref{eq:omega-update} using \eqref{eq:branch-scores}\;
        
		Update $q_\beta^{(r+1)}$ by \eqref{eq:beta-update}\;

		Fix $q_x^{(r)},\hat{\boldsymbol{\omega}}_{i,k}^{(r+1)}$,
		construct $\boldsymbol{\eta}_{R,k}^{(r)}$
		using \eqref{eq:r-conjugate-message}\;
        
		Initialize: $\tilde\bl_{k,0}^{\bR,(r)}=0$ and $\bl_{k,0}^{\bR,(r)}=\boldsymbol{\eta}_{ k}^{\bR,(r)}$\;
        
		\For{$t=1,\ldots,T_{\bR}$}{
			Update $q_{R,t}^{(r)}$ via \eqref{eq:r-update}-\eqref{eqn-3.2} with MC gradients from \eqref{eq:r-expectation-gradient}, holding $\boldsymbol{\eta}_{R,k}^{(r)}$ fixed\;
		}
		Set $q_R^{(r+1)}$ to the resulting posterior\;

		Fix $q_R^{(r+1)},q_P^{(r)},
		\hat{\boldsymbol{\omega}}_{i,k}^{(r+1)}$;
		construct $\boldsymbol{W}_{i,k}^{(r+1)}$ via \eqref{eqn-3.3}\;
        
		\For{$t'=1,\ldots,T_{\bx}$}{
			Compute \eqref{eq:x-grad-mean}-\eqref{eq:x-grad-cov}, then update $q_{x,t'}^{(r)}$ via
			\eqref{eq:x-mean-update}-\eqref{eq:x-cov-update}\;
		}

        Assign the final state iterate to $q_x^{(r+1)}$\;
        
		Update $q_P^{(r+1)}$ via \eqref{eq:p-update}\;

		\If{Condition \eqref{eq:stopping-rule}  is satisfied}{
			\textbf{break}\;
		}
	}
	\Return\;
\end{algorithm}

	\subsection{Discussion: degenerate to CVIAKF when $\rho_k=\beta_k=1$}
	
In this subsection, we shall show that CVIAKF-DR reduces to CVIAKF in \cite{10818708}, if $\rho_k=\beta_k=1$.
	
	\begin{proposition}
		\label{lem:tac-reduction}
		With no packet dropouts ($\rho_k=1$) and no corruption
		(either $\beta_k=1$ or $\boldsymbol{E}_k=0$), CVIAKF-DR reduces to CVIAKF.
	\end{proposition}
\begin{proof}
	Without packet dropout and corruption, we have
	$\mathcal I_k=\{1,\ldots,L\}$ and $\hat\omega^{(r)}_{i,k}=1$. The only parameters need to be estimate are $\bTheta_k:=(\bx_k,\bP_{k|k-1},\bR_k)$ as we introduced in Section \ref{sec-2.3}.
    \begin{enumerate}
        \item Inference of $\bR_k$: from \eqref{eq:r-exact-coordinate} only the conjugate part $\tilde p_{c,k}^{(r)}(\bR_k,\by_{i,k})$ remains, since $\hat\omega_{i,k}^{(r)}=1$. Due to the natural parameter $\boldsymbol{\eta}_k^{\bR,(r)}$ in \eqref{eq:r-conjugate-message}, we have
    \begin{align*}
		\hat{v}_{k|k}^{(r+1)}&=\hat{v}_{k|k-1}+L,\notag\\
		\boldsymbol{V}_{k|k}^{(r+1)}&=\boldsymbol{V}_{k|k-1}+\sum_{i=1}^{L}\E_{q_x^{(r)}}\!\left[
		(\boldsymbol{y}_{i,k}-\boldsymbol{h}_k(\boldsymbol{x}_k))(\cdot)^T\right].
	\end{align*}
    \item Inference of $\bx_k$: from \eqref{eqn-3.3} the effective precision becomes
	\begin{align}\notag
		\boldsymbol{W}_{i,k}^{(r+1)}=&\E_{q_R^{(r+1)}}[\boldsymbol{R}_k^{-1}]\\
		=&(\hat{v}_{k|k}^{(r+1)}-d_y-1)\left(\boldsymbol{V}_{k|k}^{(r+1)}\right)^{-1}.
		\label{eq:reduced-state-precision}
	\end{align}
	Hence \eqref{eq:x-cov-update}-\eqref{eq:x-mean-update} recover the
	state inference in CVIAKF, Theorem 2, \cite{10818708}.
        \item Inference of $\bP_{k|k-1}$: The conjugate
	$\boldsymbol{P}_{k|k-1}$ update is unchanged in \eqref{eq:p-update}.
    \end{enumerate}
    \end{proof}

	\section{Numerical experiments}\label{sec:experiments}

    We adopt the nonlinear re-entry tracking scenario S5 in Section IV of \cite{10818708} as the benchmark to compare our method with existing ones. The state
	$\boldsymbol{x}_k=[x_k,y_k,\dot{x}_k,\dot{y}_k,u_k]^{ T}$
	contains position, velocity, and the aerodynamic property. The nonlinear dynamic model is
	\begin{equation}
		\boldsymbol{f}(\boldsymbol{x}_k)=
		\begin{bmatrix}
			x_k+\Delta t\dot{x}_k\\
			y_k+\Delta t\dot{y}_k\\
			\dot{x}_k+\Delta t(D_k\dot{x}_k+G_kx_k)\\
			\dot{y}_k+\Delta t(D_k\dot{y}_k+G_ky_k)\\
			0
		\end{bmatrix},
		\label{eq:s5-f}
	\end{equation}
	where $\Delta t=0.1$, the drag-related force 
	\begin{align*}
		D_k&=\gamma_0\exp(u_k)
		\exp[(R_e-M_k)/H_0]V_k,
	\end{align*}
    and the gravity-related force
    \begin{align*}
		G_k&=-G_{m0}/M_k^3.
	\end{align*}
	Here, $M_k=\sqrt{x_k^2+y_k^2}$ and
	$V_k=\sqrt{\dot{x}_k^2+\dot{y}_k^2}$ are the radial
	distance and speed, respectively. The remaining
	parameters are $\gamma_0=-0.59783$, $H_0=13.406$,
	$G_{m0}=3.986\times10^5$, and the Earth's radius $R_e=6374$. The units in this experiment are kilometer (km) and second (s). The initial state is
	$\boldsymbol{x}_0=[6500.4,349.14,-1.8093,-6.7969,0.6932]^{ T}$ and the initial covariance is
	$\boldsymbol{P}_0=\operatorname{diag}(100,100,100,100,1)\times 10^{-4}$. Each MC run draw initial states $\bx_0$ from
	$\mathcal N(\boldsymbol{x}_0,\boldsymbol{P}_0)$.
	The nominal process covariance is
	$\boldsymbol{Q}_0=\operatorname{diag}(1,1,2.406,2.406,0.1)\times 10^{-4}$.	
	
	The observations consist of range and azimuth:
	\begin{equation}
		\boldsymbol{h}(\boldsymbol{x}_k)=
		\begin{bmatrix}
			\sqrt{(x_k-s_x)^2+(y_k-s_y)^2}\\
			\arctan\!\left(y_k-s_y)/(x_k-s_x)\right)
		\end{bmatrix},
		\label{eq:s5-h}
	\end{equation}
	where $(s_x,s_y)=(0,R_e)$ is the radar location.
	The nominal measurement covariance is
	$\boldsymbol{R}_0=\operatorname{diag}(100,
	1)\times 10^{-6}$. The true noise covariances and known corruption increment are
	\begin{equation}
		\begin{aligned}
			\boldsymbol{Q}_k=&\left(1+\tfrac12\cos(\pi k/N)\right)\boldsymbol{Q}_0,\\
			\boldsymbol{R}_k=&\left(1+\tfrac12\cos(\pi k/N)\right)\boldsymbol{R}_0,\\
			\qquad \boldsymbol{E}_k=&100\boldsymbol{R}_0.
		\end{aligned}
		\label{eq:s5-noise}
	\end{equation}
	
The total experimental time is $T=300$s. We perform $M=5$ MC runs, each with 	$K=\frac{T}{\Delta t}=\frac{300}{0.01}=3000$ steps and $L=200$ conditionally independent sensor observations per step.
The dropout probability \(1-\rho_k\) rises from \(0.20\) to \(0.50\) for \(100\leq t<200\,\mathrm{s}\), and the conditional corruption probability \(1-\beta_k\) increases from \(0.15\) to \(0.40\). Both revert to their original values thereafter.
	
The parameters  \eqref{eq:iwr-param-19} of the AKF in Section \ref{sec-2.3} is set to be $\widehat{\boldsymbol{Q}}_k=10\boldsymbol{Q}_0$,
	$\widehat{\boldsymbol{R}}_0=0.1\boldsymbol{R}_0$, and $\tau_P=50$.
	The covariance forgetting factor $\varrho_R$ is $0.95$;
	both probability forgetting factors in \eqref{eq:beta-forgetting-rho} are $\varrho_\rho=\varrho_\beta=0.60$.
	The step sizes of CVI in \eqref{eq:r-update} and \eqref{eq:x-cov-update}-\eqref{eq:x-mean-update} are $\alpha_{\bx,t'}=\alpha_{\bR,t}=0.2$, the stopping
	tolerance in \eqref{eq:stopping-rule} is $10^{-7}$.

	The compared methods are listed below:
	\begin{enumerate}
		\item CVIAKF-DR: the proposed adaptive filter with
		packet-dropout and corruption inference.
          \item G-GIWM-VB-CKF \cite{Yun2022RobustCKF}: robust VB-CKF with normal measurement noise covariance and outlier noise covariance.
		\item CVIAKF \cite{10818708}: CVI-based joint state and noise estimation.
		 \item VBAKF-CM: the extension of VBAKF\cite{Huang2018AdaptiveKF} for nonlinear
models using unbiased measurement conversion\cite{6850165}.
         \item UKFNCM: UKF using the nominal process covariance $\boldsymbol{Q}_0$ and the nominal measurement covariance $\boldsymbol{R}_0$.
      	\item UKFTCM: UKF using the true $\boldsymbol{Q}_k$ and $\boldsymbol{R}_k$.	 
	\end{enumerate}
	
The six filters form three pairs to enable multi-level comparisons. First, CVIAKF-DR and G-GIWM-VB-CKF are robust adaptive filters for unreliable measurements: the former jointly infers packet dropouts and measurement corruption using our shared-baseline covariance model, while the latter models nominal and outlier noise covariances separately to handle outliers. Second, CVIAKF and VBAKF-CM are adaptive-noise baselines without our dual-reliability mechanism: VBAKF-CM leverages standard VB with inverse-Wishart covariance adaptation, and CVIAKF applies CVI for nonlinear-induced nonconjugate inference. Finally, UKFNCM and UKFTCM are nonadaptive references with identical NLF structures but distinct noise information: UKFNCM uses fixed nominal covariances, while UKFTCM adopts true time-varying covariances as an oracle benchmark quantifying noise-statistics uncertainty effects.

	Performance is evaluated using RMSE and its
	time average, denoted by ARMSE.
	For example, the RMSE of position is 
\begin{equation}
		\operatorname{RMSE}_{\rm pos}(k)=
		\sqrt{\frac{1}{M}\sum_{j=1}^{M}
			\left[
			(\widehat{x}_k^{(j)}-x_k^{(j)})^2+
			(\widehat{y}_k^{(j)}-y_k^{(j)})^2
			\right]}.
		\label{eq:s5-rmse}
	\end{equation}

    	\begin{figure}[h!]
		\centering
		\includegraphics[width=\linewidth]
		{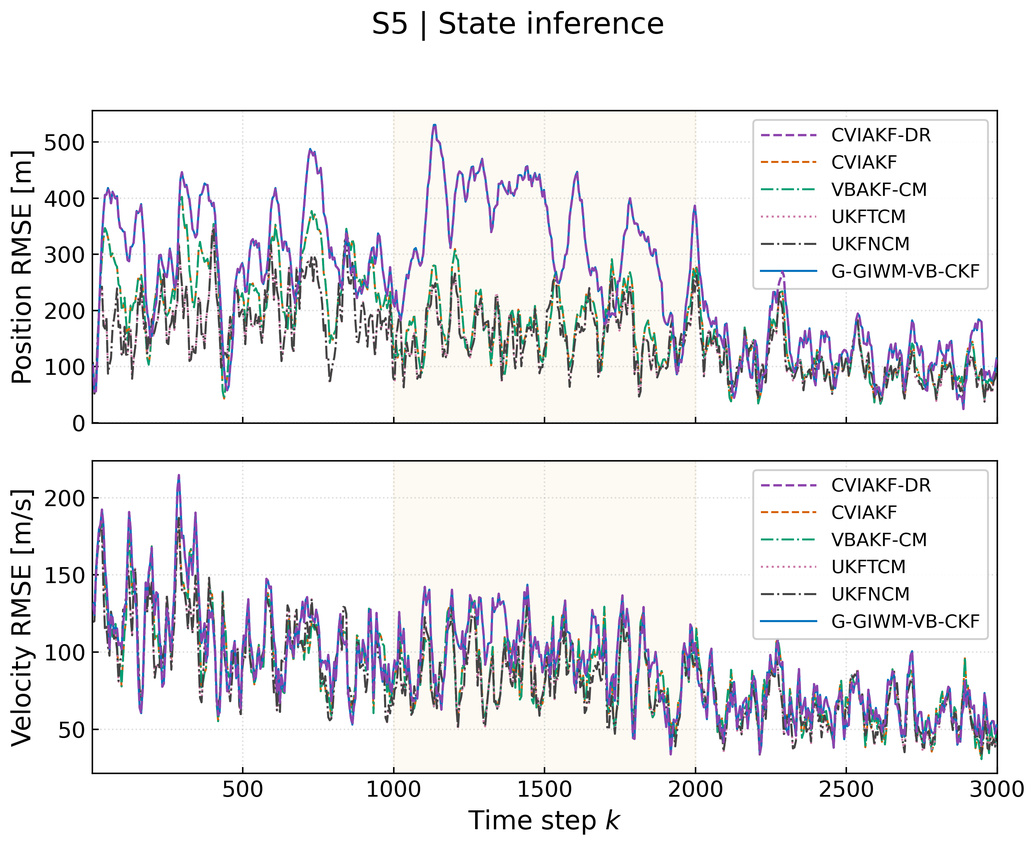}
		\caption{Position and velocity RMSEs in S5.}
		\label{fig:s5-state}
	\end{figure}
	
		\begin{table}[h!]
		\centering
		\caption{State ARMSEs in S5}
		\label{tab:s5-state}
		\setlength{\tabcolsep}{4pt}
		\begin{tabular}{lcc}
			\toprule
			Method & Position (m) & Velocity (m/s)\\
			\midrule
			CVIAKF-DR    & {\bf 257.742} & {\bf 93.301}\\
            G-GIWM-VB-CKF$^{1}$ & 257.953 & 93.302\\\hline
			CVIAKF        & 175.115 &  87.266\\
			VBAKF-CM      & 175.141 &  87.267\\\hline
			UKFNCM        & 150.284 &  83.364\\
            UKFTCM        & 150.235 &  83.289\\
			\bottomrule
		\end{tabular}
       \\[2pt]
         \footnotesize{$^1$ The G-GIWM-VB-CKF is adapted to multiple observations and observable dropouts, with an additional
	$\boldsymbol{P}_{k|k-1}$ inference using the same nominal
	$\widehat{\boldsymbol{Q}}_k=10\boldsymbol{Q}_0$ as CVIAKF-DR. The two covariance priors start from $\widehat{\boldsymbol{R}}_0$ and $\widehat{\boldsymbol{R}}_0+\boldsymbol{E}_k$.}
	\end{table}

	\begin{figure}[h!]
		\centering
		\includegraphics[width=\linewidth]
		{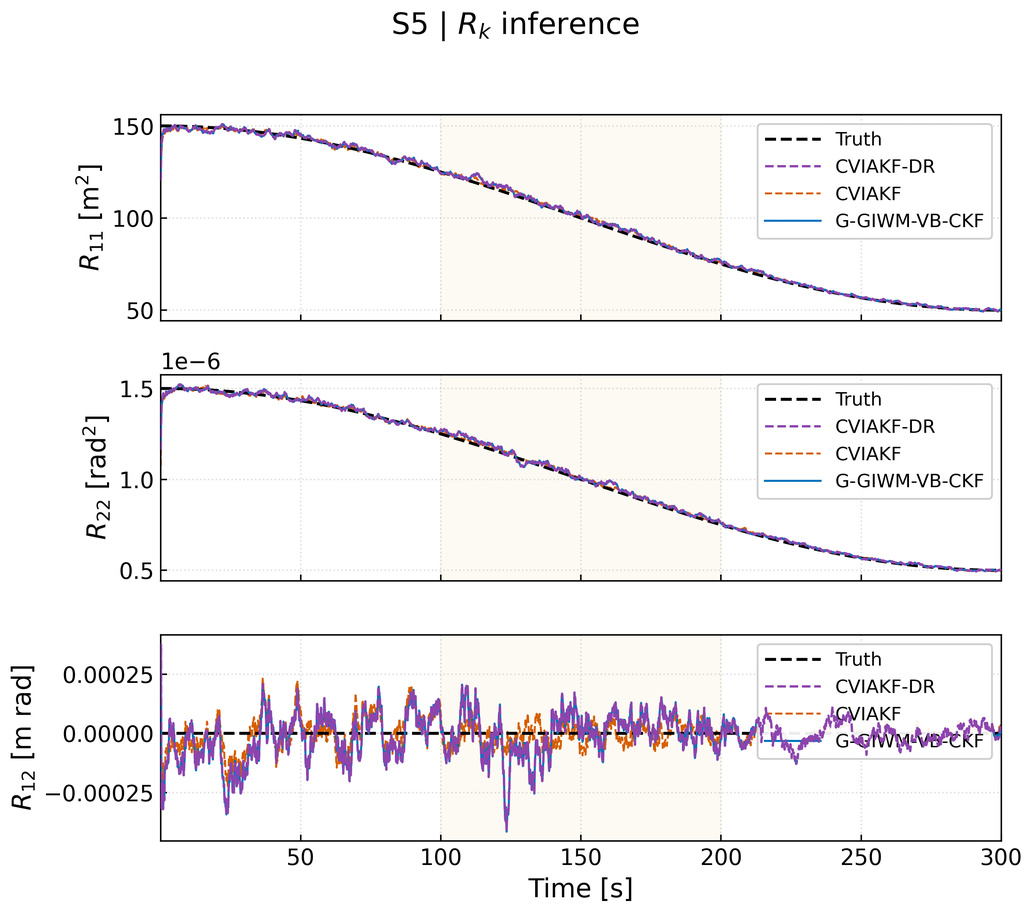}
		\caption{Estimates of $R_{11}$, $R_{22}$, and
			$R_{12}$ in S5. Curves are posterior means
			averaged over $5$ MC runs.}
		\label{fig:s5-r}
	\end{figure}
	
	\begin{table}[h!]
		\centering
		\caption{Entrywise covariance ARMSEs in S5}
		\label{tab:s5-r}
		\setlength{\tabcolsep}{3pt}
		\begin{tabular}{lccc}
			\toprule
			Method & $R_{11}\,(\mathrm{m}^2)$ & $R_{22}\,(10^{-8}\,\mathrm{rad}^2)$ & $R_{12}\,(10^{-4}\,\mathrm{m}\cdot\mathrm{rad})$\\
			\hline
			CVIAKF-DR     & {\bf 2.30696} & {\bf 2.26180} & {\bf 1.53256}\\
            G-GIWM-VB-CKF & 2.31947 & 2.26892 & 1.53407\\\hline
			CVIAKF        & 1.69516 & 1.67601 & 1.12319\\
			\bottomrule
		\end{tabular}
	\end{table}

	\begin{figure}[h!]
		\centering
		\includegraphics[width=\linewidth]{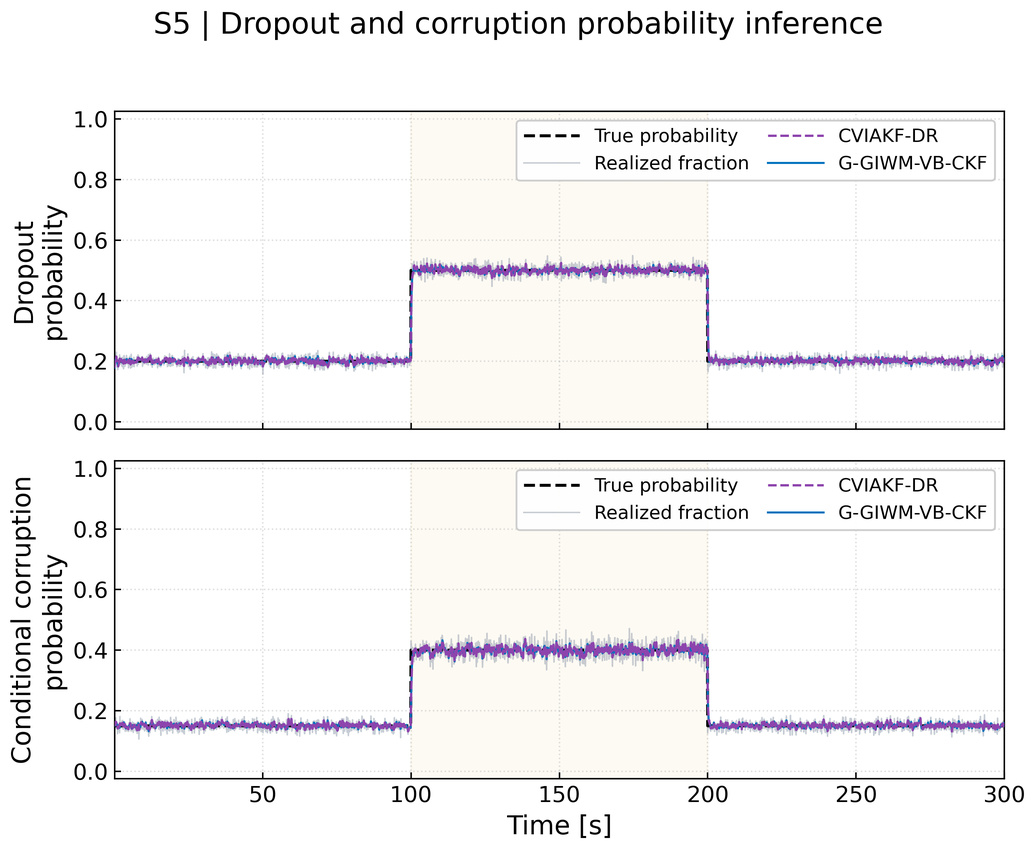}
		\caption{Dropout and conditional corruption
			probabilities in S5, averaged over $5$ MC runs.}
		\label{fig:s5-rates}
	\end{figure}
	
Fig.~\ref{fig:s5-state} and Table~\ref{tab:s5-state} compare the state-estimation results. 
CVIAKF-DR performs slightly better than G-GIWM-VB-CKF, achieving lower ARMSEs in both position and velocity.

Fig.~\ref{fig:s5-r} shows that all three methods accurately track $\boldsymbol{R}_k$, even during intervals with increased dropouts and corruption. Table~\ref{tab:s5-r} shows CVIAKF-DR achieves slightly smaller covariance ARMSEs than G-GIWM-VB-CKF, but larger errors than CVIAKF (which uses complete and correct observations). This is because CVIAKF-DR exploits the known structure $\boldsymbol{R}_k+\boldsymbol{E}_k$, whereas G-GIWM-VB-CKF separately estimates the two branch covariances. Well-separated noise levels and multiple sensor observations allow both methods to distinguish the branches, yielding similar accuracy.

As shown in Fig.~\ref{fig:s5-rates}, both methods track the probability changes at $100$ and
	$200\,\mathrm{s}$ with slight delays. Both methods achieve a dropout ARMSE of $0.01469$. Regarding the corruption rate, the ARMSEs are $0.016915$ for CVIAKF-DR and $0.016911$ for G-GIWM-VB-CKF, indicating essentially comparable reliability-estimation accuracy.
	\section{Conclusion}\label{sec:conclusion}
This work develops CVIAKF-DR to jointly estimate state and noise covariance with observable packet reception and unknown measurement authenticity. Adopting the shared covariance structure $\boldsymbol{R}_k$ and $\boldsymbol{R}_k+\boldsymbol{E}_k$, the method leverages CVI for state and covariance updates while maintaining closed-form updates for prediction covariance and reliability probabilities. We rigorously prove that CVIAKF-DR degenerates to CVIAKF without packet dropouts and corruption, and simulations confirm its ability to track time-varying covariance and abrupt variations in dropout and corruption probabilities.

Future work will integrate nonlinear filtering methods such as CKF, UKF and particle filtering into the CVI framework to address constrained covariance inference and problems with missing or corrupted measurements, while evaluating estimation accuracy, convergence and computational overhead.

	\bibliographystyle{ieeetr}
	\bibliography{references1}
	
\end{document}